# "This (Smart) Town Ain't Big Enough": Smart Small Towns and Digital Twins for Sustainable Urban and Regional Development

Gabriela Viale Pereira[1], Lukas Daniel Klausner[2], Lucy Temple[1], Thomas Delissen[2], Thomas Lampoltshammer[1] and Torsten Priebe[2]

[1] *University for Continuing Education Krems, Krems, Austria*
[2] *St. Pölten University of Applied Sciences, St. Pölten, Austria*

**Abstract**
One of the major challenges today lies in the creation of governance concepts for regional development that not only promote growth but, at the same time, ensure promotion of inclusiveness, fairness, and resilience. Digital twins can support policymakers in developing smart, sustainable solutions for cities and regions and, therefore, urban and non-urban environments. The project SCiNDTiLA (Smart Cities aNd Digital Twins in Lower Austria) aims to define the state-of-the-art in the field of smart cities, identify interdependencies, critical components and stakeholders, and provide a roadmap for smart cities with application to both smaller-scale urban and non-urban environments. SCiNDTiLA uses the foundations of complexity theory and computational social science methods to model Austrian towns and regions as smart cities/regions and thus as systems of socio-technical interaction to guide policy decision-making toward sustainable development.

## 1. Introduction

This paper gives a brief introduction to the project **S**mart **Ci**ties a**N**d **D**igital **T**wins **i**n **L**ower **A**ustria (SCiNDTiLA, 2023–2026), in which we take stock of existing research on smart cities and digital twins and reshape and develop the state of the art for a new context. To wit, thus far smart city concepts have almost exclusively been implemented in the context of large urban environment, such as Barcelona, London, Milan and Seoul [1–3]. In contrast, SCiNDTiLA will complement existing research by adapting smart city and digital twin approaches to small-town and (urban–)rural regional sprawl use cases, with a particular focus on the interaction of policymaking and sustainable local governance and the needs, desires and expectations of the towns' and/or regions' inhabitants.

The main research question we have identified is: How can the existing knowledge on smart cities be transferred to smaller-scale urban and non-urban contexts, and how can technologies such as digital twins be used to support policymakers in developing smart sustainable solutions for cities and regions in Lower Austria?

Consequently, our project's objectives are:

1. defining the state of the art in the field of smart city with focus on the sustainable aspects, including governance and social, economical and environmental dimensions, and identifying the characteristics that can be transferred to smaller-scale urban and non-urban contexts based on a systematic literature review and best practices assessment;

2. developing a conceptual framework of sustainable local governance via digital twins to validate the smart city generic model through an innovative transdisciplinary process, including the requirements elicitation for use cases in small cities in Lower Austria;
3. developing a digital-twin-based sustainable smart city and defining different scenarios concerning challenges of good governance in smaller-scale urban and non-urban contexts; and finally;
4. implementing the proof-of-concept use cases in Lower Austria and proposing a roadmap highlighting methodologies, guidelines, and policy recommendations on how smart and sustainable solutions in cities and regions shape inhabitants' perceptions of local governance.

Our approach thus combines a systemic foundation, identifying interdependencies, critical components and fundamental stakeholders, and an operational upscaling to provide a roadmap for smart city and digital twin applications in both smaller-scale urban and non-urban environments. Moreover, our project team brings together an interdisciplinary team from both social and technical sciences and from various career stages.

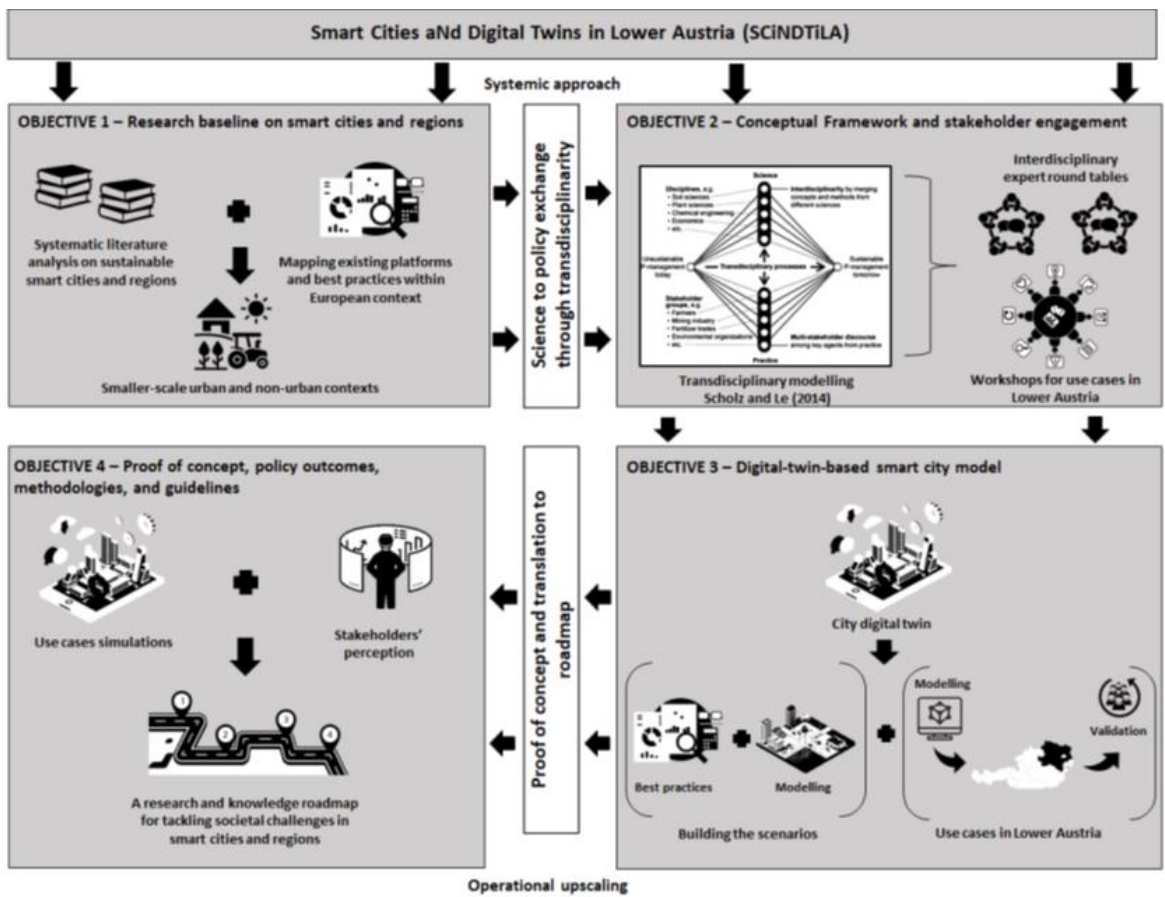

**Figure 1**: Schematic overview over SCiNDTiLA's objectives and planned schedule

As illustrated in Figure 1, the four objectives of the project provide a comprehensive response to the research question. The outcomes will establish the link between policy development towards sustainable smart cities/regions and their impacts on the public perceptions of local governance as well as the way government tackles societal challenges. As the two research agendas have thus far been developed in relative isolation, this joint approach can make a unique contribution and improve the effectiveness of smart city initiatives implemented in Lower Austria and beyond.

This project is applied in the context of Lower Austria, which is characterised by the growing number of policies that have been developed to support "digitalisation of the public sector" and, in particular, so-called "smart initiatives" at the regional and local level according to the Digitalization Strategy of Lower Austria. Regarding sustainable impact, a well-formulated data governance strategy is imperative towards evidence-based policymaking. Yet this governance strategy alone is not sufficient: To ensure the improvement of the overall well-being of inhabitants, participation and communication during the development, execution and evaluation of policies are key in the context of the big societal challenges, i. e. climate change, energy and beyond.

## 2. Related Work and Main Concepts

There are many ways to approach the concept of governance in the context of smart cities. Existing research on smart cities with a governance focus is mostly seen from a user-centred perspective, with more emphasis on citizens and other stakeholders and the idea of developing productive interactions between networks of urban actors. Smart cities were defined by Rodríguez Bolívar and Meijer [4] as cities with smart collaboration for problem-solving. More broadly, smart cities involve sharing responsibility and authority between local governments and/or governmental departments, citizens, the private sector and other stakeholders working together towards data-driven decision-making [5]. Therefore, smart-city governance requires a strong focus on expanding collaboration with the wider local community and strengthening the cooperation of different stakeholders, which can be facilitated by increased engagement with beneficiaries of services and policies [6]. Smart-city strategies have been shown to intensify collaboration amongst stakeholders, with increased dependency on the context in which the city is embedded [7].

Moreover, the project highlights the importance of an information systems (IS) focus on smart-city-related studies as most existing studies lack a detailed analysis from an IS context, especially regarding the implementation and use of IS to design and develop smart cities [8]. Therefore, we take an IS perspective to address the smart city concept, following the approach by Ismagilova et al. [8, p. 90]: "Smart cities use an IS centric approach to the intelligent use of ICT within an interactive infrastructure to provide advanced and innovative services to its citizens, impacting quality of life and sustainable management of natural resources." This perspective is aligned with SCiNDTiLA's approach, especially regarding the way citizens and other stakeholders contribute to creating public value (e. g. collaboration, innovation, transparency, sustainable development) through the use of open data to better serve citizens' needs [9], using smart technologies to support citizen-driven decision-making [10] and using ICT in the collaborative decision-making process in smart cities [5]. Viale Pereira et al. [11] describe smart governance as the intelligent use of ICT to improve decision-making and, particularly related to smart-city governance, state that it fundamentally deals with government decisions for improving quality of life and emphasises the citizens' role in collaborative decision-making.

Complementing the smart city approach, digital twin technology has also seen increasing interest in both academia and industry, offering many benefits ranging from cost reduction in manufacturing [12] to supporting decision-making in asset lifecycle management processes [13]. Although different definitions of the digital twin concept exist, we will follow the definition of VanDerHorn and Mahadevan [14], who define a digital twin as "a virtual representation of a physical system (and its associated environment and processes) that is updated through the exchange of information between the physical and virtual systems". In our project, we focus on the application of digital twins to smart cities as described by Wang et al. [15]; in their work, most of the smart city research projects focus on large, densely populated urban areas, where data are already collected on a large scale. We expect that creating digital twins for small towns and rural areas will pose additional challenges, but also offer additional opportunities – in particular because employing digital twins allows for context-specific aspects to be considered and deepens understanding of the kinds of conditions required for certain policies to succeed [16]. The use of digital twins has helped identify problems such as mismatches between residents' needs, a lack of facilities or inadequate use of infrastructure [17]. This may enable more effective policy-making and a more evidence-based approach. The creation of a simulated environment begins with the identification of those elements that comprise the system, the boundaries of the digital twin and the rules that will define its behaviour and the behaviour of its elements. While working in small towns and rural areas, it is important to clearly identify the resident population and which elements they seek to include or exclude from the system, thus heightening the value of involving relevant stakeholders who thus act as a collaborative ecosystem that facilitates innovation [18].

## 3. Methodology

To address the interplay between people, organisations and technology as the nature of information systems research, the research method of this project follows a design science research approach [19] to define a smart city as a system (IT artifact) to be evaluated in a given organisational context – in this project, the state of Lower Austria. The conceptualisation of a smart city as an IT artifact includes the definition of constructs (vocabulary and symbols), models (abstractions and representations), methods (algorithms and practices) and instantiations (implemented and prototype systems) [19], allowing us to understand the problems (most pressing challenges of local governance in sustainable smart cities) and the feasibility of possible solutions (collaborative/evidence/algorithm-based decision-making processes of the policymakers towards smart solutions for regions, cities, and communities).

The project will also apply a novel transdisciplinary process to conceptualise sustainable smart cities and validate the generic model, including contextualising and identifying the main societal challenges to be addressed as use cases in small cities and regions in Lower Austria. Transdisciplinary processes organise societal learning processes to develop socially robust orientations for sustainable development [20]. "Socially robust" means that the orientation includes scientific state-of-the-art knowledge, the results from integrating and relating scientific knowledge (i. e. academic rigour) and best knowledge from science (i. e. experiential wisdom), whereby not only the uncertainty, but also the ignorance of human knowledge are acknowledged, and the constraints of the study become transparent [21]. To apply this approach, we initiate the process by an interdisciplinary roundtable and follow up on this with a moderated multi-stakeholder discourse to identify the most pressing challenges for local governance in sustainable smart cities. During the course of the project, we will devise a methodology for assessing policies and their effectiveness according to the changes in the state of the system and not just according to the intended or desired effects of those policies. With the use of algorithmic decision support and digital twins, our project will deliver a major step towards using simulated environments within which both artificial agents and humans (via their digital twins) can interact with each other and thereby improve the quality of conclusions and learnings from employing this kind of technology.

## 4. Expected Results

Due to its interdisciplinary nature, the expected results of the project will have implications in a variety of areas. We expect this to include

- contributions on modelling smart city ecosystems and adapting them to the specific context of Lower Austria, including for small towns, rural areas, regions and communities;
- the use of digital twins and other simulation and forecasting techniques in areas other than smart cities, such as the industrial sector [22], and specifically closing the simulation-to-reality gap therein [23];
- empirical contributions in the understanding of how digital applications and smart city initiatives shape sustainable behaviour at the city government level [5];
- conceptualisation of sustainable smart cities and future research trends with emphasis on their sustainable and societal impact [24, 25];
- contributions to interdisciplinary socio-technological understandings at the intersection of smart city/digital twins and science and technology studies, focusing on the mutual impacts of digital-twin-based simulations in smaller towns and non-urban environments and public perceptions of technology and government [26];
- methodological contributions in how transdisciplinary processes can shape complex debates on the effects of digitisation in smart cities [20, 27, 28];
- critical reflections on what the underlying implicit and explicit assumptions inherent to simulations, digital twins and smart cities/regions mean for policy-makers and society at large [29];
- explore the applicability of algorithmic decision support to assist policymakers in developing smart sustainable solutions [30].

The main contribution to practice consists in the translation of the results into a roadmap to guide policy outcomes and allow for generalisable findings for smart cities and local-government-driven projects in new environments and use cases. We will develop a research and knowledge roadmap for tackling societal challenges in smart cities with a focus on rescaling the entire framework to be transferred to small towns and non-urban environments, such as rural areas and smart villages.

## 5. Acknowledgements

This research was funded by the Gesellschaft für Forschungsförderung Niederösterreich (GFF NÖ) project GLF21-2-010 "Smart Cities and Digital Twins in Lower Austria". The financial support by the Gesellschaft für Forschungsförderung Niederösterreich is gratefully acknowledged.

## 6. References

[1] Simon Elias Bibri and John Krogstie. The Emerging Data-Driven Smart City and Its Innovative Applied Solutions for Sustainability: The Cases of London and Barcelona. Energy Inform., 3, 2020.
[2] Mila Gascó, Benedetta Trivellato, and Dario Cavenago. How Do Southern European Cities Foster Innovation? Lessons from the Experience of the Smart City Approaches of Barcelona and Milan. In J. Ramon Gil-Garcia, Theresa A. Pardo, and Taewoo Nam, editors, Smarter as the New Urban Agenda: A Comprehensive View of the 21st Century City, Public Administration and Information Technology, vol. 11, pages 191–206. Springer, Cham, 2016.
[3] Jung-Hoon Lee, Marguerite Gong Hancock, and Mei-Chih Hu. Towards an Effective Framework for Building Smart Cities: Lessons from Seoul and San Francisco. Technol. Forecast. Soc. Change, 89:80–99, 2014.
[4] Manuel Pedro Rodríguez Bolívar and Albert J. Meijer. Smart Governance: Using a Literature Review and Empirical Analysis to Build a Research Model. Soc. Sci. Comput. Rev., 34(6):673–692, 2016.
[5] Gabriela Viale Pereira, Maria Alexandra Cunha, Thomas J. Lampoltshammer, Peter Parycek, and Maurício Gregianin Testa. Increasing Collaboration and Participation in Smart City Governance: A Cross-Case Analysis of Smart City Initiatives. Inf. Technol. Dev., 23(3):526–553, 2017.
[6] Gianluca Misuraca, Fiorenza Lipparini, and Giulio Pasi. Towards Smart Governance: Insights from Assessing ICT-Enabled Social Innovation in Europe. In Elsa Estevez, Theresa A. Pardo, and Hans Jochen Scholl, editors, Smart Cities and Smart Governance: Towards the 22nd Century Sustainable City, Public Administration and Information Technology, vol. 37, pages 217–238. Springer, Cham, 2021.
[7] Jessica Clement, Miguel Manjon, and Nathalie Crutzen. Factors for Collaboration Amongst Smart City Stakeholders: A Local Government Perspective. Gov. Inf. Q., 39(4), 2022.
[8] Elvira Ismagilova, Laurie Hughes, Yogesh K. Dwivedi, and K. Ravi Raman. Smart Cities: Advances in Research—An Information Systems Perspective. Int. J. Inf. Manag., 47:88–100, 2019.
[9] Gabriela Viale Pereira, Marie Anne Macadar, Edimara M. Luciano, and Maurício Gregianin Testa. Delivering Public Value Through Open Government Data Initiatives in a Smart City Context. Inf. Syst. Front., 19(2):213–229, 2017.
[10] Gabriela Viale Pereira, Gregor Eibl, Constantinos Stylianou, Gilberto Martínez, Haris Neophytou, and Peter Parycek. The Role of Smart Technologies to Support Citizen Engagement and Decision Making: The SmartGov Case. Int. J. Electron. Gov. Res., 14(4):1–17, 2018.
[11] Gabriela Viale Pereira, Peter Parycek, Enzo Falco, and Reinout Kleinhans. Smart Governance in the Context of Smart Cities: A Literature Review. Inf. Polity, 23(2):143–162, 2018.
[12] Liwen Hu, Ngoc-Tu Nguyen, Wenjin Tao, Ming C. Leu, Xiaoqing Frank Liu, Md Rakib Shahriar, and S. M. Nahian Al Sunny. Modeling of Cloud-Based Digital Twins for Smart Manufacturing with MT Connect. Procedia Manuf., 26:1193–1203, 2018.
[13] Marco Macchi, Irene Roda, Elisa Negri, and Luca Fumagalli. Exploring the Role of Digital Twin for Asset Lifecycle Management. IFAC-PapersOnLine, 51(11):790–795, 2018.

[14] Eric VanDerHorn and Sankaran Mahadevan. Digital Twin: Generalization, Characterization and Implementation. Decis. Support Syst., 145, 2021.
[15] Hao Wang, Xiaowei Chen, Fu Jia, and Xiaojuan Cheng. Digital Twin-Supported Smart City: Status, Challenges and Future Research Directions. Expert Syst. Appl., 217, 2023.
[16] Negar Noori, Thomas Hoppe, Martin De Jong, and Evert Stamhuis. Transplanting Good Practices in Smart City Development: A Step-Wise Approach. Gov. Inf. Q., 40(2), 2023.
[17] Qing Geng and YingGe Du. From Blockchain to Digital Twin Community: A Technical Framework for Smart Community Governance. In 2021 International Conference on Public Management and Intelligent Society, PMIS 2021, pages 277–280, Washington, DC, 2021. IEEE.
[18] Francesco Paolo Appio, Marcos Lima, and Sotirios Paroutis. Understanding Smart Cities: Innovation Ecosystems, Technological Advancements, and Societal Challenges. Technol. Forecast. Soc. Change, 142:1–14, 2019.
[19] Alan R. Hevner, Salvatore T. March, Jinsoo Park, and Sudha Ram. Design Science in Information Systems Research. Manag. Inf. Syst. Q., 28(1):75–105, 2004.
[20] Roland W. Scholz and Gerald Steiner. The Real Type and Ideal Type of Transdisciplinary Processes: Part I—Theoretical Foundations. Sustain. Sci., 10(4):527–544, 2015.
[21] Roland W. Scholz. Environmental Literacy in Science and Society: From Knowledge to Decisions. Cambridge University Press, Cambridge, 2011.
[22] Jamilya Nurgazina, Thomas Felberbauer, Bernward Asprion, and Pavan Pinnamaraju. Visualization and Clustering for Rolling Forecast Quality Verification: A Case Study in the Automotive Industry. Procedia Comput. Sci., 200:1048–1057, 2022.
[23] Oliver Eigner, Sebastian Eresheim, Peter Kieseberg, Lukas Daniel Klausner, Martin Pirker, Torsten Priebe, Simon Tjoa, Fiammetta Marulli, and Francesco Mercaldo. Towards Resilient Artificial Intelligence: Survey and Research Issues. In 2021 IEEE International Conference on Cyber Security and Resilience, CSR 2021, pages 536–542, Washington, DC, 2021. IEEE.
[24] Luiza Schuch de Azambuja, Gabriela Viale Pereira, and Robert Krimmer. Clearing the Existing Fog over the Smart Sustainable City Concept: Highlighting the Importance of Governance. In Proceedings of the 13th International Conference on Theory and Practice of Electronic Governance, ICEGOV '20, New York, NY, 2020. ACM.
[25] Gabriela Viale Pereira and Luiza Schuch de Azambuja. Smart Sustainable City Roadmap as a Tool for Addressing Sustainability Challenges and Building Governance Capacity. Sustainability, 14(1), 2022.
[26] Angelika Adensamer, Rita Gsenger, and Lukas Daniel Klausner. "Computer Says No": Algorithmic Decision Support and Organisational Responsibility. J. Responsible Technol., 7–8, 2021.
[27] Roland W. Scholz, Eric J. Bartelsman, Sarah Diefenbach, Lude H. Franke, Armin Grunwald, Dirk Helbing, Richard Hill, Lorenz Hilty, Mattias Höjer, Stefan Klauser, Christian Montag, Peter Parycek, Jan-Philipp Prote, Ortwin Renn, André Reichel, Günther Schuh, Gerald Steiner, and Gabriela Viale Pereira. Unintended Side Effects of the Digital Transition: European Scientists' Messages from a Proposition-Based Expert Round Table. Sustainability, 10(6), 2018.
[28] Gabriela Viale Pereira, Elsa Estevez, Diego Fernando Cardona Madarriaga, Carlos I. Chesñevar, Pablo Collazzo-Yelpo, Maria Alexandra Cunha, Eduardo Henrique Diniz, Alex Antônio Ferraresi, Frida Marina Fischer, Flúvio Cardinelle Oliveira Garcia, Luiz Antonio Joia, Edimara M. Luciano, João Porto de Albuquerque, Carlos O. Quandt, Rodrigo Sánchez Rios, Aurora Sánchez Ortiz, Eduardo Damião da Silva, João Silvestre Silva-Junior, and Roland W. Scholz. South American Expert Roundtable: Increasing Adaptive Governance Capacity for Coping with Unintended Side Effects of Digital Transformation. Sustainability, 12(2), 2020.
[29] Angelika Adensamer and Lukas Daniel Klausner. "Part Man, Part Machine, All Cop": Automation in Policing. Front. Artif. Intell., 4, 2021.
[30] Gabriele De Luca, Thomas J. Lampoltshammer, and Shahanaz Parven. Why Developing Simulation Capabilities Promotes Sustainable Adaptation to Climate Change. In Artificial Intelligence in HCI, HCII 2021/Lecture Notes in Artificial Intelligence, vol. 12797, pages 490–500, Cham, 2021. Springer.